\documentclass{emulateapj}
\usepackage{epsfig} 
\usepackage{natbib}
\usepackage{epstopdf}
\usepackage{graphicx}
\usepackage{epstopdf}

\newcommand{\sub}[2]{\ensuremath{#1_{\mathrm{#2}}}}

\newcommand{\unit}[2]{\ensuremath{\textrm{#1}^{#2}}}

\newcommand{\name}{ULAS~J001535.72$+$015549.6}
\newcommand{\nametwo}{ULAS~J074417.48$+$253233.0}

\newcommand{\namesh}{ULAS~J0015$+$01}
\newcommand{\nametwosh}{ULAS~J0744$+$25}

\newcommand{\paperI}{\citetalias{2014AJ....147...76B}}
\defcitealias{2014AJ....147...76B}{Paper~I}

\shortauthors{Bochanski et al.} \shorttitle{The Most Distant Stars in the Galaxy} 
\begin{document}
\slugcomment{Accepted by ApJL}

\title{The Most Distant Stars in the Milky Way}

\author{ John J. Bochanski\altaffilmark{1}, Beth Willman\altaffilmark{1}, Nelson Caldwell\altaffilmark{2}, Robyn Sanderson\altaffilmark{3}, Andrew A. West\altaffilmark{4}, Jay Strader\altaffilmark{5}, Warren Brown\altaffilmark{2}}

\altaffiltext{1}{Haverford College, 370 Lancaster Ave, Haverford PA 19041 USA\\
email:jbochans@haverford.edu} 
\altaffiltext{2}{Harvard-Smithsonian Center for Astrophysics, Cambridge, MA, 02138}
\altaffiltext{3}{Kapteyn Astronomical Institute, P.O. Box 800, 9700 AV Groningen, The Netherlands}
\altaffiltext{4}{Department of Astronomy, Boston University, 725 Commonwealth Avenue, Boston, MA 02215 USA} 
\altaffiltext{5}{Michigan State Astronomy Group, Michigan State University, Biomedical Physical Sciences Building, 567 Wilson Road, Room 3261 East Lansing, MI  48824-2320 USA}

\begin{abstract}
We report on the discovery of the most distant Milky Way (MW) stars known to date: {\name} and {\nametwo}. These stars were selected as M giant candidates based on their infrared and optical colors and lack of proper motions.  We spectroscopically confirmed them as outer halo giants using the MMT/Red Channel spectrograph.  Both stars have large estimated distances, with {\name} at $274 \pm 74$ kpc and {\nametwo} at 238 $\pm$ 64 kpc, making them the first MW stars discovered beyond 200 kpc.  {\name} and {\nametwo} are both moving away from the Galactic center at $52 \pm 10$ km s$^{-1}$ and $24 \pm 10$ km s$^{-1}$, respectively.  Using their distances and kinematics, we considered possible origins such as: tidal stripping from a dwarf galaxy,  ejection from the MW's disk, or membership in an undetected dwarf galaxy.  These M giants, along with two inner halo giants that were also confirmed during this campaign, are the first to map largely unexplored regions of our Galaxy's outer halo.

\end{abstract}

\section{Introduction} 
The outer halo of our Milky Way (MW) has yet to be comprehensively mapped.  Surveys, such as the Sloan Digital Sky Survey \citep[SDSS;][]{2000AJ....120.1579Y} have yielded large samples of metal-poor main sequence turnoff stars, which have been used to map out the MW's inner halo \citep[e.g.,][]{bell08a} to distances of $d \lesssim$ 50 kpc over large areas of sky.  Other stellar tracers, such as RR Lyrae stars \citep[e.g.,][]{2013ApJ...765..154D}, red giant branch stars \citep[e.g.,][]{helmi03a,2014ApJ...784..170X} and blue horizontal branch (BHB) stars \citep[e.g.,][]{schlaufman09a} have also been used to map the Galactic halo to distances out to $d \sim 120$ kpc.  However, very few stars at $d \gtrsim 120$ kpc have been identified mostly due to the faint limits of large surveys ($r \lesssim 21$).  Table \ref{table:distant_stars} lists the known stars at $d > 120$ kpc.  

This severely incomplete picture of the Galaxy's outermost reaches limits our understanding of halo formation and evolution.  Stars are not expected to be able to form \textit{in situ} at outer halo distances \citep[e.g.,][]{2009ApJ...702.1058Z}.  
Instead, lone stars found at truly large distances ($d \gtrsim 200$ kpc) could mark the presence of outer halo substructure from accreted dwarf galaxies \citep[e.g.,][]{2005ApJ...635..931B,2003ApJ...599.1082M}, or result from a star being ejected from the Galactic center or disk \citep[e.g.,][]{2005ApJ...622L..33B}.  If they are bound to the MW, distant halo stars could be used to measure the MW's virial mass, which is uncertain by a factor of $\sim 3$, ranging from $\sim 0.7$ to $2 \times10^{12} M_\odot$ \citep{Sofue2012,deason12a,Burch2013,Irrgang2013}.

M giants currently offer the best opportunity to probe the MW's outer halo with existing surveys.  They are intrinsically bright, with typical bolometric luminosities of $\log L/L_{\odot} \sim 3-4 $.  
Given their relatively cool effective temperatures, near infrared (NIR) surveys, such as the Two--Micron All Sky Survey \citep[2MASS;][]{2006AJ....131.1163S} and the UKIRT Infrared Deep Sky Survey \citep[UKIDSS;][]{2007MNRAS.379.1599L} are especially sensitive to M giants.  M giants in 2MASS were used to extensively map out the Sagittarius dwarf galaxy \citep[Sgr;][]{2003ApJ...599.1082M,2010ApJ...714..229L}.  SDSS observations of M giants were also used to map out the Sgr tidal remnants, including the structure at distances near $\sim 100$ kpc \citep{2009ApJ...700.1282Y,2014MNRAS.437..116B}.  The UKIDSS data, which has a faint limit 3--4 mag deeper than 2MASS, is sensitive to the most luminous M giants to distances beyond the MW's virial radius \citep[$\sim$ 180-300 kpc,][and references therein]{2002ApJ...573..597K, 2012MNRAS.425.2840D, 2014A&A...562A..91P}.  

In \citealp{2014AJ....147...76B} (hereafter Paper~I), we assembled a catalog of 404 M giant candidates, which were selected based on their NIR colors from the UKIDSS Large Area Survey (ULAS), optical colors from SDSS, and proper motions.  In this Letter, we report on the discovery of two extremely distant M giants from this catalog, {\name} and {\nametwo} (hereafter {\namesh} and {\nametwosh}).  These stars were spectroscopically confirmed as M giants, at distances of $\sim$ 270 and 240 kpc (and at least 180 and 130 kpc, respectively, depending on metallicity).  They are the first MW stars identified beyond 200 kpc, and potentially beyond the virial radius.  
We present our observations in Section \ref{sec:obs}.  Our spectral type determinations, distance estimates and radial velocity measurements are described in Section \ref{sec:analysis}.  In Section \ref{sec:results} we discuss our hypotheses for the origins of these stars.

\begin{deluxetable*}{rrrrlcr}
\tablewidth{0pt}
\tabletypesize{\small}
 \tablecaption{Distant Stars in the Milky Way}
 \tablehead{
 \colhead{R.A. (deg)} &
 \colhead{Dec. (deg)} &
 \colhead{$\ell$ (deg)} &
 \colhead{$b$ (deg)} &
  \colhead{Distance (kpc)} &
  \colhead{Type} &
 \colhead{Reference} 
}
 \startdata
115.65748  & 22.97206   &  197.11 &	21.07 & 122 $\pm$ 12\tablenotemark{a} &RRL &  \cite{2013ApJ...765..154D}\\ 
  261.4764 & 3.0072& 25.62 &	20.36& 126 $\pm$ 32\tablenotemark{b} &  CN & \cite{2008AA...482..151M}\\
 22.4696 & 3.2119 &141.41 &	-58.275& 133 $\pm$ 6 & BHB & \cite{2012MNRAS.425.2840D} \\
  341.6206 & -27.4501&24.98	&-62.31& 145 $\pm$ 36\tablenotemark{b} & CN & \cite{2005AA...438..867M}\\
4.4821 & 0.0631 &105.07&	-61.64&  151 $\pm$ 7 & BHB & \cite{2012MNRAS.425.2840D} \\
  136.4432& 20.4106& 207.52&	38.35&153 $\pm$ 38\tablenotemark{b}&   CN & \cite{2012MNRAS.425.2840D}\\
  195.3269 & 0.4975&308.42&	63.26& 161 $\pm$ 40\tablenotemark{b} &   CN & \cite{2008AA...482..151M}\\
\hline
 116.07282 & 25.54249 &194.68&	22.34& 238 $\pm$ 64 & M giant & This Paper   \\                    
3.89882 & 1.93044 & 104.97&	-59.69&274 $\pm$ 74 & M giant & This Paper

\enddata
 \label{table:distant_stars}
 \tablecomments{We present stars with estimated distances $>$ 120 kpc.}
\tablenotetext{a}{Assuming a 10\% uncertainty in distance, as explained in \cite{2013ApJ...765..154D}.}
\tablenotetext{b}{Following \cite{2012MNRAS.425.2840D}, we present the mean distance estimated using the \cite{2000MNRAS.314..630T} and \cite{2004AA...418...77M} methods, and assume an uncertainty of 25\%.}
\end{deluxetable*}

\section{Observations}\label{sec:obs}
The details of our M giant candidate selection are contained in {\paperI}, but briefly provided here.  We selected M giants using NIR color cuts, similar to the \cite{2003ApJ...599.1082M} and \cite{2010ApJ...722..750S} color cuts, but shifted to reduce contamination from K/M dwarfs and K giants.  
Our NIR targets were matched to  SDSS observations, which were used to discriminate against quasars. Reddening is small throughout the sample.  For example, $E(B-V)$ values for {\namesh} and {\nametwosh} are 0.03 and 0.04, respectively.  We astrometrically screened out M dwarfs using the SDSS-USNOB proper motion catalogs \citep{2004AJ....127.3034M}, supplemented by proper motions derived from SDSS-UKIDSS astrometry.  After all cuts were applied, 404 M giant candidates remained.  Our initial study netted five M giants, and our spectroscopic campaign is ongoing.

During 2013 Nov 12--14, we obtained spectra of 32 stars using the Red Channel Spectrograph \citep[RCS;][]{1989PASP..101..713S} at the 6.5m MMT observatory at Mount Hopkins, Arizona.  We used the RCS in single order mode, employing the 1.0$^{\prime\prime}$ slit, 1200 lines mm$^{-1}$ grating centered at 8400 \AA\ and LP-530 long-pass filter, resulting in a resolution of $R \sim 5000$. Due to variable and significant cloud coverage during the first night, we observed relatively bright ($V < 10$) stars previously classified as M giants and M dwarfs to aid in spectroscopic classification.  These stars, which serve as high signal--to--noise (S/N) spectral standards, are listed in Table \ref{table:observed}.  Over the remaining two nights, we observed 15 M giant candidates, listed in Table \ref{table:observed}.  Science observations ranged from 180s to 7200s of total exposure time, broken into multiple exposures to aid in cosmic ray rejection.  The data were reduced using an IDL software package based on the MASE \citep{mase} reduction pipeline.  Each single-order observation was bias-corrected, flat-fielded, wavelength calibrated, and optimally extracted.  Wavelength calibrations were obtained using the HeNeAr arc lamp, and corrected to the heliocentric rest frame.  Flux calibration was computed by comparing to standards, with at least one standard being observed per night.  The typical S/N of our science observations ranged from $\sim$ 10 to 50.  The spectra of many of the stars obtained during our run are shown in Figure \ref{fig:templates}.

\begin{deluxetable*}{rrrclrl}
\tablewidth{0pt}
\tabletypesize{\small}
 \tablecaption{Observed M Giant Candidates \& Spectral Standards}
 \tablehead{
 \colhead{Name} &
 \colhead{R.A. (deg)} &
 \colhead{Dec. (deg)} &
 \colhead{$J$} &
 \colhead{Sp.Type} &
 \colhead{RV\tablenotemark{a}  (km s$^{-1}$)}&
 \colhead{Notes} 
}
 \startdata
ULAS J001535.72+015549.6  &   3.89882  &   1.93045 & 17.73 &  M giant & -58 $\pm$ 10  & M giant at $\sim$ 274 kpc. \\
ULAS J021121.56$-$003808.5  &  32.83984  &  -0.63568 & 15.58 & M dwarf & &  \\
ULAS J073223.56+263420.0  & 113.09816  &  26.57222 & 16.49 &  M dwarf & & \\
ULAS J074150.40+263355.5  & 115.46001  &  26.56542 & 13.88 & M dwarf & &  \\
ULAS J074417.48+253233.0  & 116.07282  &  25.54249 & 17.28 &  M giant & 86 $\pm$ 10 & M giant at $\sim$ 238 kpc. \\
ULAS J075202.79+204645.0  & 118.01162  &  20.77916 & 13.41 &  \nodata & & Inconclusive spectrum \\
ULAS J075525.09+235952.2  & 118.85454  &  23.99783 & 14.64 &  M giant & 99 $\pm$ 10 & M giant at $\sim$ 50 kpc. \\
ULAS J075554.26+273130.9  & 118.97609  &  27.52525 & 15.23 & M giant & 17 $\pm$ 10 & M giant at $\sim$ 52 kpc ({\paperI})\\
ULAS J205800.46$-$003445.3  & 314.50190  &  -0.57925 & 17.63 &  M dwarf & &  \\
ULAS J211225.54$-$005329.4  & 318.10643  &  -0.89151 & 18.17 &  M dwarf & &  \\
ULAS J223815.77+042531.9  & 339.56569  &   4.42554 & 17.31 &  M dwarf & & \\
ULAS J225509.66+114543.6  & 343.79025  &  11.76211 & 14.12 & M giant & -260 $\pm$ 10&  M giant at $\sim$ 35 kpc. \\
ULAS J225630.15+065544.8  & 344.12564  &   6.92911 & 17.83 &  M dwarf & & \\
ULAS J231331.68+065303.0  & 348.38200  &   6.88417 & 18.10 &  M dwarf & & \\
ULAS J235247.04+023151.5  & 358.19599  &   2.53098 & 17.14 & M dwarf & & \\
\hline
HD 4647 & 12.30779 & 57.07502 & 3.52 & M2 III & & Standard  \\
WW Psc & 14.95703 & 6.48322 & 2.65 & M2 III & &  Standard \\
V360 And & 16.01874 & 38.68854 & 3.38 & M3 III & &  Standard \\
HD 236791 &23.71879& 59.41716 & 4.97 & M3 III & &  Standard \\
BD+18 372 &	 43.65878  & 19.34401 & 6.12 & M3 III & & Standard \\
RZ Ari & 43.95208 & 18.33164 & 0.22 & M6 III & &  Standard \\
HD 17993 & 44.123918 &62.60961 & 3.44 & M1 III & &  Standard \\
EH Cet & 44.26905 & 4.50102 & 2.18 & M4 III & &  Standard \\
SS Cep & 57.37506 & 80.32247 & 0.76 & M5 III & &  Standard \\
HD 24410 & 58.99011 &57.67356 & 3.02 & M8 III & &  Standard \\
GX And& 4.59536 & 44.02295  & 5.25 & M2 V &&   Standard \\
GQ And& 4.60624 & 44.02712 & 6.79 & M6 V & &  Standard \\
V596 Cas& 29.84798 & 58.52113 & 7.79 & M4 V & &  Standard \\
GJ 3136 & 32.22332 & 49.44906 & 8.42 & M5 V &&  Standard  \\
HD 15285& 36.94109 & 4.43215 & 5.99 & M1 V & &  Standard \\
HIP 20745& 66.67843 & 12.68658 & 7.82 & M0 V & &  Standard \\
YZ Cmi& 116.16739 & 3.55245 & 6.58 & M5 V & &  Standard

\enddata
 \label{table:observed}
 \tablenotetext{a}{Radial velocities in the heliocentric rest frame.}
\end{deluxetable*}

\section{Analysis}\label{sec:analysis}

\subsection{Spectral Type Estimates}
Spectral types for our science targets were estimated using two methods.  First, each science spectrum was compared to the 17 giant and dwarf spectra.  We visually compared each standard-science spectral pair, and computed the $\chi^2$ residuals over the wavelength regime shown in Figure \ref{fig:templates}.  The spectra of {\namesh} and {\nametwosh} agreed more closely with the giant standards, both visually and with respect to $\chi^2$ residuals.  M dwarf and M giant template spectra were also constructed by coadding individual giant and dwarf spectra.  Each standard observation (gray lines in Figure \ref{fig:templates}) was normalized with a fourth-order polynomial to remove the stellar continuum prior to coaddition.  The coadded template spectra are shown in Figure \ref{fig:templates}, along with the spectra of {\namesh} and {\nametwosh}.  The dominant atomic spectral features in this wavelength regime are the Na I doublet near 8200 \AA\ and the Ca II triplet near 8500 \AA.  TiO gives rise to the bandheads near 8400 \AA, which are easily observed in the M giant spectra, as they change strength drastically throughout the sequence.  Telluric features are notable as well.  The absorption seen at 8227 \AA\ is due to terrestrial water vapor \citep{1974ApJ...189..463A}.  The emission features near 8345, 8400, 8430, 8505, and 8780 \AA\ are due telluric OH emission lines \citep{2006JGRA..11112307C}. 

In Figure \ref{fig:compare}, we compare the spectra of the distant M giants to the M giant (top) and M dwarf (bottom) templates.  The M giant template is a better match to both stars.  While the S/N in the {\namesh} spectrum is only $\sim$ 10, the Na I doublet near 8200 \AA\ is not visible.   The spectrum of {\nametwosh} has a higher S/N ($\sim$ 20), and is well matched by the M giant template.   While M dwarfs exhibit Ca II absorption, the M giant template is a better match on the blue side of the spectrum, and displays no Na I absorption.  Furthermore, if either star was an M dwarf, the Na I doublet would be of similar strength to the water vapor near 8227 \AA.  However, there are no strong absorption bands in this regime.  We note that low--mass subdwarfs also have weakened Na I absorption.  Recently, Savcheva et al. (submitted) prepared a catalog of 3517 low--mass subdwarfs with SDSS spectra.  Less than 1\% of these stars would have passed our color and proper motion cuts.  We visually inspected the SDSS spectra of subdwarfs that did, and all displayed prominent Na I absorption.  Thus, {\namesh} and {\nametwosh} were classified as M giants.  The M giant with the closest match to both stars, both visually and with respect to $\chi^2$ was EH Cet, an M4 III giant with an $M_J = -4.6$ \citep{2007A&A...474..653V}. 

\subsection{Distance Estimates}
As noted in {\paperI}, precise photometric distance estimates of M giants are notoriously difficult to produce.  Depending on the relation assumed, the absolute magnitude is independent of color \citep[i.e.,][]{2000ApJ...542..804N,2009ApJ...700.1282Y}, relates linearly with $J-K$ color \citep{2010ApJ...722..750S} or can be estimated using stellar evolution models \citep{2012ApJ...759..131B}. For each star, we computed distances with six different techniques: either assuming [Fe/H]$=0.0, -0.5, -1.0$ and employing the \cite{2012ApJ...759..131B} method, using the absolute magnitude relations from \cite{2010ApJ...722..750S}, \cite{2000ApJ...542..804N}, \cite{2009ApJ...700.1282Y}, and \cite{2012AJ....143..128P}, or assuming the $M_J$ of EH Cet.  The distance estimates are contained in Table \ref{table:distances}.  For each distance estimate, we assumed an uncertainty of 25\%, which is similar to the uncertainty assumed for carbon giants \citep{2005AA...438..867M,2012MNRAS.425.2840D}.

Six of the ten estimates indicate a distance between 260 and 290 kpc for {\namesh}.  For {\nametwosh}, seven of the ten distance estimates fall between 210 and 290 kpc.  For both stars, the assumed [Fe/H] has a large effect on the distance estimate.  For all of the metallicities used in this analysis, the distance to {\namesh} is $\gtrsim$ 180 kpc, making it the most distant MW star known to date.   Both stars are best matched to the spectrum of EH Cet.  The absolute magnitude of EH Cet yields distances of $290 \pm 73$ and $238 \pm 59$ for {\namesh} and {\nametwosh}, respectively.  These values happen to be close to the mean values from all of the distance estimates.  Therefore, we adopt the mean distance for each star with the standard deviation of all measurements as the uncertainty and report them in Tables \ref{table:distant_stars} and \ref{table:distances}.

\subsection{Radial Velocity Measurements}
We measured the radial velocity of each M giant by cross-correlating the spectrum against SDSS spectra of M giants.  In {\paperI}, we recovered four M giants in the SDSS database, with typical uncertainties of 10 km s$^{-1}$.  Given the limited wavelength range sampled by MMT, we sought to minimize the effect of exactly which region was selected for cross-correlation.  Each science target was correlated against each SDSS M giant spectrum 1000 times, slightly adjusting the starting and end points of the region used for correlation each time.  Telluric regions were masked out during the RV measurement.  The mean radial velocity and standard deviation was recorded for each SDSS star.  The mean heliocentric radial velocity of {\namesh}, measured against all four SDSS M giants, was $-57 \pm 10$ km s$^{-1}$.  The measured velocity of {\nametwosh} was  $86 \pm 10$ km s$^{-1}$.  We converted these velocities using the circular speed (240 km s$^{-1}$) and solar motion used in \cite{2012MNRAS.425.2840D}, resulting in Galactocentric velocities of $52 \pm 10$ km s$^{-1}$ for {\namesh} and $24 \pm 10$ km s$^{-1}$ for {\nametwosh}.  These velocities are consistent with the relatively cold velocity dispersions seen at large distances \citep{2012MNRAS.425.2840D}. We discuss the implication of these velocities on the origins of these stars in the following section.

\begin{deluxetable*}{lccll}
\tablewidth{0pt}
\tabletypesize{\small}
 \tablecaption{Distance Estimates of {\namesh} and {\nametwosh}}
 \tablehead{
 \colhead{Method} &
 \colhead{Description} &
  \colhead{Assumed [Fe/H]} &
 \colhead{$d$ (kpc)\tablenotemark{a}} &
  \colhead{$d$ (kpc)\tablenotemark{b}} 
}
 \startdata
\cite{2000ApJ...542..804N}  & $M_K = -5.5$   & $\sim -0.7$ & $276 \pm 69$  & $224 \pm 56$ \\
\cite{2009ApJ...700.1282Y}  & $M_g = -1.0$ & $\sim -0.8$  & $271 \pm 68$ & $294 \pm 74$   \\
\cite{2009ApJ...700.1282Y}  & $M_r \sim -2.3$ & $\sim -0.8$  & $262 \pm 66$  & $264 \pm 66$ \\
\cite{2010ApJ...722..750S}  & $M_K = 3.26 - 9.42\times(J-K)$   & $ \gtrsim -1$  & $415 \pm 104$ & $347 \pm 87$\\
\cite{2012AJ....143..128P}  & $M_i = -1.5$   & $\sim 0$  & $133 \pm 33$ & $128 \pm 32$\\
\cite{2012AJ....143..128P}  & $M_z = -3.5$   & $\sim 0$  & $281 \pm 70$ & $265 \pm 66$\\
\cite{2012ApJ...759..131B}  & $P(M_H | J-K)$   & 0.0 & $178\pm 45$ & $134\pm 34$ \\
\cite{2012ApJ...759..131B}  & $P(M_H | J-K)$   & -0.5 & $281\pm 70$ & $214\pm 53$\\
\cite{2012ApJ...759..131B}  & $P(M_H | J-K)$   & -1.0 & $354 \pm 89$ & $274 \pm 68$\\
Comparison to EH Cet  & $M_J \sim -4.6$     & \nodata & $290 \pm 73$ & $238 \pm 59$ \\
\hline
Adopted Distance & Mean &  \nodata & $274 \pm 74$ & 238 $\pm$ 64
\enddata
\label{table:distances}
 \tablenotetext{a}{Distance estimates for {\namesh}.}
 \tablenotetext{b}{Distance estimates for {\nametwosh}.}
\end{deluxetable*}

\section{Results and Conclusions}\label{sec:results}
The origins of {\namesh} and {\nametwosh} are interesting, given their large distances from the MW.  They are at least three times as distant as the Large Magellanic Cloud, and potentially more distant than Leo I and Leo II  \citep{1950PASP...62..118H}.  Both stars are near the MW's virial radius, where gas densities are extremely low, and star formation is virtually non-existent.  Thus, it would be overwhelmingly unlikely that either star formed \textit{in situ}.

Given the largely accepted accretion model for the formation of the MW's halo, the most natural hypothesis for the origin of these stars is accretion from a  dwarf galaxy \citep[e.g.,][]{2005ApJ...635..931B,2009ApJ...702.1058Z}.  Such tidal stripping is well mapped in the inner halo, which is dominated by the Sgr dwarf and its well-mapped tidal tails \citep{2003ApJ...599.1082M,2014MNRAS.437..116B}.  However, only one stream has been identified at $d$ $>$ 100 kpc \citep{2013ApJ...765..154D}.   We compared the position and radial velocities of both stars to a model of Sgr \citep{2010ApJ...714..229L}. Unlike the closer M giants associated with Sgr in {\paperI}, we did not definitively associate these M giants with Sgr.  However, we do note that both stars lie close to the Sgr plane, with {\nametwosh} at $(\Lambda,B)_{\rm Sgr} \sim (190.2, 4.7)$, and {\namesh} at $(\Lambda,B)_{\rm Sgr} \sim (81.8, -16.2)$.  Thus, an association with Sgr cannot be ruled out.  Furthermore, {\namesh} is spatially coincident with the apocenter of the Sgr trailing arm, while {\nametwosh} is spatially coincident with the Pisces over-density \citep{2010ApJ...722..750S}.  Despite being within $\sim$ 10$^{\circ}$ of these structures, both stars have much larger distances.

Despite differences in formation history and halo masses, simulations predict that the majority of stars at $d > 150$ kpc have been accreted less than 9 Gyr ago, suggesting recent accretion is important in the outer halo \citep{2009ApJ...702.1058Z}.  Models also predict significantly less substructure with $d \gtrsim 160$ kpc and a stellar density over four orders of magnitude smaller than the solar radius \citep{2005ApJ...635..931B}.  While the predicted amount of substructure in the outer halo is small, our experiment was designed to target any trace of this population.  Therefore, it is possible that {\namesh} and {\nametwosh} have been stripped from substructure accreted in the last 10 Gyr.  With more complete maps and UKIDSS sky coverage, this hypothesis can be more fully explored. 

An alternative hypothesis for these stars' origin is ejection from the MW.  Both star--binary (i.e., disk ejection) and star--black hole (i.e., hypervelocity) interactions can eject stars \citep[i.e.,][]{2013ApJ...768..153Z,2012ApJ...751..133P}.  To test this mechanism, we calculated the time it took each star to travel to its current position and velocity, assuming it was formed near the solar circle and was ejected.  
Since the halo is the dominant potential at these distances, we used a spherical NFW halo \citep{1996ApJ...462..563N} to model the Galactic potential:
\begin{equation}
\Phi(r) = -\frac{G \sub{M}{vir}}{a f(c)} \frac{\ln(1+r/a)}{r/a},
\end{equation}
where the scale radius $a$ is related to the concentration $c$ of the halo via $a\equiv \sub{r}{vir}/c$, and the concentration constant $f(c)$ is
\begin{equation}
f(c) \equiv \ln(1+c) - \frac{c}{1+c}.
\end{equation}
Recent estimates of the concentration parameter $c$, range from $\sim$ 18 to 24 \citep{Battaglia2005,deason12a,Smith2007}. As mentioned above, the current estimated for the virial mass of the MW range from $\sim 0.7$ to $2 \times10^{12} M_\odot$ \citep{Sofue2012,deason12a,Burch2013,Irrgang2013}.  Thus, we adopted the middle values of the MW's mass ($1.4\times10^{12}\ M_\odot$) and concentration parameter ($c=20$) as fiducial values for the MW potential.

First, we calculated the speeds with which the stars would have left the disk, along with the corresponding apocenters. Assuming each star started near the solar circle (8 kpc), the initial Galactocentric velocities were similar: $\sub{v}{r,J0015+01}=572$ km \unit{s}{-1} and $\sub{v}{r,J0744+25}=563$ km \unit{s}{-1}. The additional potential terms of the disk and bulge were ignored, but they would serve to increase these speeds. Even without these Galactic components, the necessary speed for each star to reach its current distance is comparable to the measured escape velocity of the MW at the solar circle \citep[500-600 km \unit{s}{-1};][]{Smith2007}.  This suggests that the stars were unlikely to be ejected, as the mechanisms that can produce these speeds are relatively rare.  Hypervelocity star ejection rates are $\sim$ 1$\times 10^{-4}$ yr$^{-1}$ \citep[i.e.,][]{2013ApJ...768..153Z}, while the rate of 600 km \unit{s}{-1} disk runaway ejections is $\sim$ 1$\times 10^{-6}$ yr$^{-1}$  \citep[i.e.,][]{2012ApJ...751..133P}.

Next, we calculated the time needed to reach each stars' current position.  Using the fiducial model, it would take both stars $\sim$ 1.5 Gyr to reach their current position, while moving outward.  We varied the mass and concentration of the Halo to compute a range of travel times, which were the same for both stars and varied from $\sim 1$ to $3$ Gyr.  We will directly test the ejection hypothesis as our sample grows.  If ejection is important in placing M giants at these distances, there should be more stars seen in the direction of rotation, since they would gain a $\sim$ 240 km s$^{-1}$ boost from the MW's rotation.

Another hypothesis for the origin of these stars is their membership in a previously unseen dwarf galaxy. Nearby MW dwarf spheroidals (Draco, Ursa Minor, Sculptor, Carina, Fornax) do not host obvious M giant populations as cataloged by 2MASS with our NIR cuts applied. This paucity of 2MASS M giants in MW dwarfs is not surprising, given 2MASS's faint limit and the low metallicities of the dwarfs.  However, Carina \citep[Fe/H $= -1.72$;][]{2012AJ....144....4M} and Fornax \citep[Fe/H $= -0.99$;][]{2012AJ....144....4M} each have one to two candidate M giant stars in 2MASS at the correct photometric distances for membership.  Assuming that a galaxy capable of hosting an M giant has a mean [Fe/H] $\gtrsim -1.0$, we used the stellar mass-metallicity relation to estimate a $M_{V} \lesssim -13$ for any unseen dwarf galaxy that could host these stars \citep{2013ApJ...779..102K}.  For such a dwarf to have escaped detection, it would have to be extremely low surface brightness ($\mu_{V,}$ $\gtrsim$ 30 mag arcsec$^{-2}$).  Although unseen, the existence of such ghostly objects has been predicted \citep{2010ApJ...717.1043B, 2011ApJ...741...18B}.  To test this hypothesis, we constructed the SDSS $g,g-r$ color magnitude diagram (CMD) within 15 arcmin of each star. These CMDs were binned and subtracted with a control CMD from a nearby field of the same area.  Visual inspection of the subtracted CMDs
revealed no obvious detection of any associated red giant branches or Sgr turnoff stars near either M giant.  Multi-object spectroscopy of surrounding stars with $r > 20$ mag could provide a test this hypothesis, along with deep search for RR Lyrae stars near each M giant.  

After observing 15 M giant candidates, our spectroscopic campaign confirmed four new M giants in the MW's halo.  This selection efficiency is in-line with our initial study ({$\sim$ 20\%, {\paperI}) and we expect that $\sim$ 70 M giants will be recovered from our sample.   By identifying more of these stars spectroscopically, we hope to further refine our selection criteria.  This will be important for identifying M giants in the next generation of large surveys, such as \textit{Gaia} and LSST \citep{2001A&A...369..339P,2008arXiv0805.2366I}.  While these stars will be too faint for reliable parallax measurements with \textit{Gaia}, they may be able to recover proper motions in the halo.  For example, an M giant at \textit{Gaia}'s faint limit ($G = 20$), with a distance of 200 kpc and a tangential velocity of 100 km s$^{-1}$ will have a proper motion of $\sim$ 100 $\mu$as yr$^{-1}$.  \textit{Gaia}'s expected proper motion uncertainty at $G = 20$ will be $\sim 50$ $\mu$as yr$^{-1}$, resulting in a $2\sigma$ measurement \citep{2012Ap&SS.341...31D}.

 \begin{figure*} 
	 
  \centering 

  \includegraphics[scale=0.6]{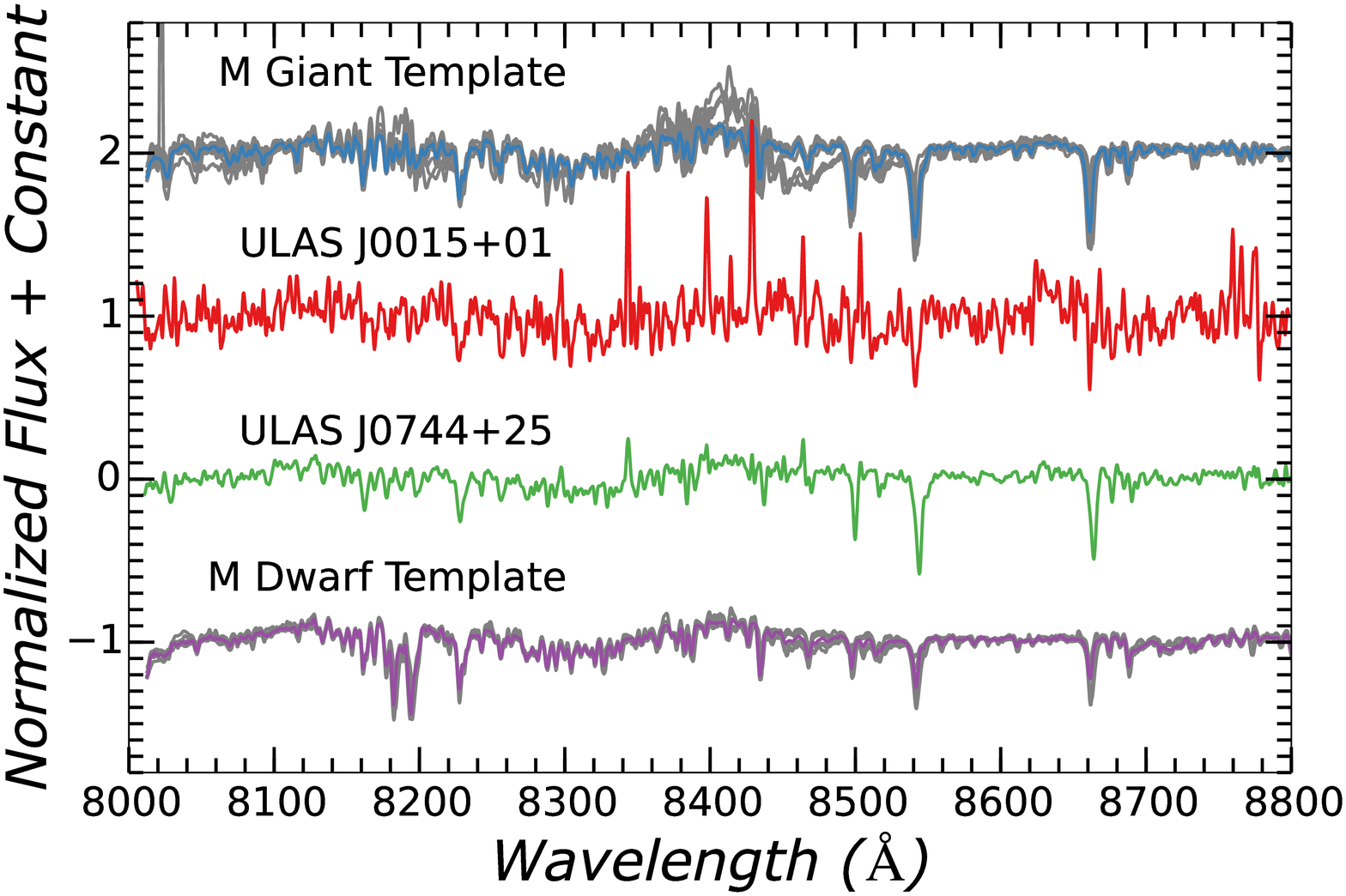} 
  \caption{The normalized standard spectra (gray lines) are shown along with the coadded M giant template (blue line) and M dwarf template (purple line).  The spectra of {\namesh} and {\nametwosh} are shown in red and green, respectively.  Note the strong Na I absorption in the M dwarfs near 8200 \AA\ and the Ca II triplet.}
  \label{fig:templates} 
 \end{figure*}

 \begin{figure*} 
  \centering 

  \includegraphics[scale=0.4]{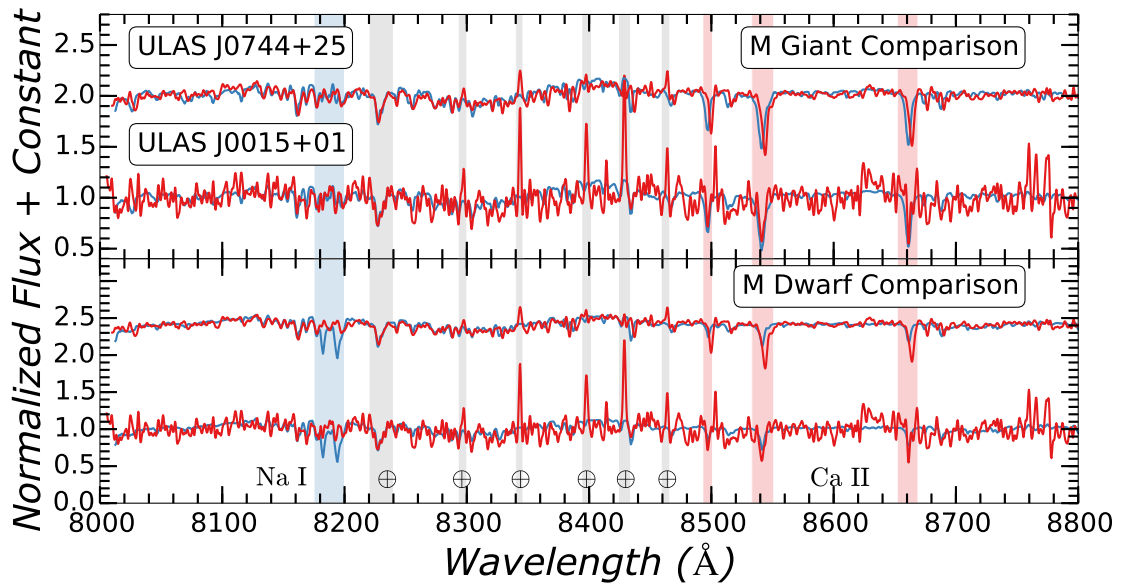} 
  \caption{The spectra of {\nametwosh} and {\namesh} (upper and lower red lines, respectively) are compared to the M giant (top panel) and M dwarf templates (bottom panel).  The Na I doublet and Ca II triplet are highlighted with blue and red shaded regions.  Telluric OH lines are marked with gray shaded regions.  The M giant template is a better match to both giants, particularly near Na I.}
  \label{fig:compare} 
 \end{figure*}

\acknowledgements 
J.J.B. and B.W. thank the NSF for support under grants NSF AST-1151462 and PHYS-1066293.  A.A.W acknowledges NSF grants AST-1109273, AST-1255568, and the RCSA's Cottrell Scholarship. We thank Jonathan Hargis, Alis Deason, Wyn Evans, Vasily Belokurov and Kathyrn Johnston for helpful conversations.

\end{document}